\documentclass{article}[12pt]
\usepackage{amsmath,amssymb}
\usepackage{graphicx}%
\usepackage{verbatim}

\newcommand{\TwoFig}[4]{%
\begin{center}
\begin{tabular}{lr}
\parbox{8cm}{\includegraphics[width=8cm]{#1}}  & \parbox{8cm}{\includegraphics[width=8cm]{#2}}  \\
\parbox{8cm}{\vspace{7pt}\refstepcounter{figure}Figure \thefigure.\quad #3\vfill} & \parbox{8cm}{\vspace{7pt}\refstepcounter{figure}Figure \thefigure.\quad #4\vfill}\\
\end{tabular}
\end{center}
\vspace{7pt}
}
\newcommand{\Fig}[3]{%
\begin{center}
\parbox{#2cm}{%
\refstepcounter{figure}\includegraphics[width=#2cm]{#1}} \parbox{12cm}{\vspace{7pt}{Figure. \thefigure.}\quad
#3}\end{center}}

\begin{document}

\begin{center}
{\bf \Large On Euclidean limit cycles in cosmological models based on scalar fields} \\[12pt]
Yu. G. Ignat'ev and A.R Samigullina\\
N. I. Lobachevsky Institute of Mathematics and Mechanics of Kazan Federal University,\\
Kremleovskaya str., 35, Kazan, 420008, Russia.
\end{center}


\begin{abstract}

A detailed analysis of the phase trajectories of cosmological models based on classical and scalar fields near surfaces of zero effective energy has been carried out.  A study of the differential parameters of the convergence of phase trajectories to a zero-energy surface boundary shows that the phase trajectories merge within a finite time with the phase trajectories of free oscillations corresponding to zero effective energy.  This confirms the assumption formulated in a number of previous works by one of the authors about the existence of Euclidean limit cycles in cosmological models based on scalar fields with a Higgs interaction potential. \\


\end{abstract}

\section*{Introduction}
In \cite{1} one of the authors\footnote{Yu.G. Ignat'ev.} proposed and partially investigated a cosmological model based on an asymmetric scalar doublet, that is, a system consisting of two scalar fields -- one, a classical field $(\Phi)$, and the other, a phantom field $(\varphi)$, with a Higgs type potential.  In \cite{2}--\cite{8} a comprehensive quantitative and numerical modeling of a cosmological model based on a classical and a phantom scalar field was carried out.  Results of these studies allowed the above-mentioned author to advance the hypothesis of the existence in such models of Euclidean limit cycles with effective zero energy, to which the system tends in the future (classical field) or in the past (phantom field).  In this setup, the Universe becomes globally Euclidean although nonzero scalar fields oscillate in it, the entire system (the Universe) finding itself in stable dynamic equilibrium.  Since the question of the existence of Euclidean limit cycles is extraordinarily important for cosmology, in the present paper we investigate this possibility in greater detail. Here, as in a number of previous papers, we will perform numerical modeling with the help of an expanded software package developed by the authors, which we call DifEqTools, specially intended to investigate nonlinear dynamical systems \cite{9}.

\section{Basic equations of a cosmological model based on an asymmetric doublet}

The Lagrange function of a scalar doublet consisting of a classical and a phantom field with self-action in Higgs form with minimal connection has the form \cite{1}
\begin{equation} \label{Eq1}
L=\frac{1}{8\pi } (g^{ik} \Phi _{,i} \Phi _{,k} -2V(\Phi ))-\frac{1}{8\pi } (g^{ik} \varphi _{,i} \varphi _{,k} +2v(\varphi )),
\end{equation}
where
\[V(\Phi )=-\frac{\alpha }{4} \left(\Phi ^{2} -e\frac{m^{2} }{\alpha } \right)^{2}, v(\varphi )=-\frac{\beta }{4} \left(\varphi ^{2} -\varepsilon \frac{{\rm m}^{2} }{\beta } \right)^{2} \]
is the Higgs potential energy of the corresponding scalar fields, $\alpha$ and $\beta$ are their self-action constants, and  \textit{m} and m are the masses of their quanta. The dynamical equations of the cosmological model based on an asymmetric scalar doublet have the following form:
\[\begin{array}{c} {\Phi '=Z,} \\ {Z'=-\sqrt{3} Z\sqrt{E_{m} (\Phi ,Z,\varphi ,z)} -e\Phi +\alpha _{m} \Phi ^{3} ,} \end{array}\]
\begin{equation} \label{Eq2}
\begin{array}{c} {\varphi '=z,} \\ {z'=-\sqrt{3} z\sqrt{E_{m} (\Phi ,Z,\varphi ,z)} +\varepsilon \mu ^{2} \varphi -\beta _{m} \varphi ^{3} ,} \end{array}
\end{equation}
where we have introduced the notation
\begin{equation} \label{Eq3}
{\mathcal E}_{m} (\Phi ,Z,\varphi ,z)=\frac{1}{2} \left[\left(Z^{2} +e\Phi ^{2} -\frac{\alpha _{m} }{2} \Phi ^{4} \right)+\left(-z^{2} +\varepsilon \mu ^{2} \varphi ^{2} -\frac{\beta _{m} }{2} \varphi ^{4} \right)+\lambda _{m} \right]
\end{equation}
for the effective energy of the scalar doublet with allowance for the cosmological constant; here $e$ and $\varepsilon =\pm 1$,
\[\lambda _{m} \equiv \frac{\lambda }{m^{2} },  \alpha _{m} \equiv \frac{\alpha }{m^{2} },  \beta _{m} \equiv \frac{\beta }{m^{2} },  \mu \equiv \frac{m}{m} .\]

In order for system of differential equations \eqref{Eq2} to have a real solution, it is necessary that the radicands be nonnegative, i.e., that the effective energy ${\rm {\mathcal E}}_{m} $ of the system with the cosmological constant taken into account be nonnegative:
\begin{equation} \label{Eq4}
{\rm {\mathcal E}}_{m} (\Phi ,Z,\varphi ,z)\ge 0.
\end{equation}
Inequality \eqref{Eq4} can lead to violation of single connection of phase space and the formation in it of closed lacunae, bounded by zero effective energy surfaces. In what follows, for brevity of the notation, we will specify the models by two ordered lists: $\mathbf{P}$ is a list of the parameters of the model and $\mathbf{I}$ is a list of the initial conditions:
\[\mathbf{P}=[\alpha _{m} ,\beta _{m} ,e,\varepsilon ,\mu ,\lambda _{m} ],\mathbf{I}=[\Phi (t_{0} ),Z(t_{0} ),\varphi (t_{0} ),z(t_{0} )].\]

\section{Zero energy trajectories}

If in Eqs. \eqref{Eq2} we set
\begin{equation} \label{Eq5}
{\rm {\mathcal E}}_{m} (\Phi ,Z,\varphi ,z)=0,
\end{equation}
the equations so obtained take the following form:
\begin{equation} \label{Eq6}
\begin{array}{c} {\Phi '=Z, Z'=-e\Phi +\alpha _{m} \Phi ^{3} ,} \\ {\varphi '=z, z'=\varepsilon \mu ^{2} \varphi -\beta _{m} \varphi ^{3} .} \end{array}
\end{equation}
These equations describe two independent dynamical systems completing anharmonic periodic oscillations in the field of a fourth-order potential.  Equations \eqref{Eq6} can be integrated exactly; however, the solution is expressed in terms of elliptic integrals and is difficult to visualize. It is important to note that the effective energy is an exact integral of these equations.  This fact enabled the author of \cite{2} to express the hypothesis that relation \eqref{Eq5} is an asymptotic exact integral of system \eqref{Eq2} for suitable values of the parameters and initial conditions.

Figures 1 and 2 show the results of numerical integration of Eqs. \eqref{Eq4}, illustrating the regime of free oscillations of the scalar doublet fields for the following values of the parameters:  $\alpha _{m} =10 beta _{m} =1 e=-1 \varepsilon =1$  and $\mu =1$.  As can be seen from these figures, the phase trajectories of system \eqref{Eq6} do indeed describe periodic anharmonic oscillations.  Note that here we did not require condition \eqref{Eq5} to be satisfied.

To start with, let us consider oscillations of a classical scalar field.  Writing out the energy of this field with the help of formula \eqref{Eq3}, we obtain
\begin{equation} \label{Eq7}
Z^{2} +e\Phi ^{2} -\frac{\alpha _{m} }{2} \Phi ^{4} =2{\mathcal E}_{0}^{c} \Rightarrow \frac{d\Phi }{d\tau } = \sqrt{2{\mathcal E}_{0}^{c} -e\Phi ^{2} +\frac{\alpha _{m} }{2} \Phi ^{4} } .
\end{equation}
Integrating Eq. \eqref{Eq7} in an interval between turning points
\begin{equation} \label{Eq8}
2{\mathcal E}_{0}^{c} -e\Phi ^{2} +\frac{\alpha _{m} }{2} \Phi ^{4} =0\Rightarrow \Phi _{k} =\sqrt{\frac{e}{\alpha _{m} } \frac{1}{|\alpha _{m} |} \sqrt{1-4{\mathcal E}_{0}^{c} \alpha _{m} } } ,
\end{equation}
we find the periods of the oscillations
\begin{equation} \label{Eq9}
T_{i} =2\int _{\Phi _{i} }^{\Phi _{i+1} }\frac{d\Phi }{\sqrt{2 {\mathcal E}_{0}^{c} -e\Phi ^{2} +{\raise0.7ex\hbox{$ 1 $}\!\mathord{\left/ {\vphantom {1 2}} \right. \kern-\nulldelimiterspace}\!\lower0.7ex\hbox{$ 2 $}} \alpha _{m} \Phi ^{4} } }  ,
\end{equation}
where the integrals are taken between two turning points (Eqs. (8)) located in the general region of accessibility.  Depending on the parameters of the field model and the total energy, solution of Eq. \eqref{Eq8} can give 0, 2, or 4 turning points. We have a similar situation for the phantom field.  Integral \eqref{Eq9} can be expressed in terms of the elliptic functions $F(\nu ,\gamma )$  and $K(\nu ,\gamma )$.

\TwoFig{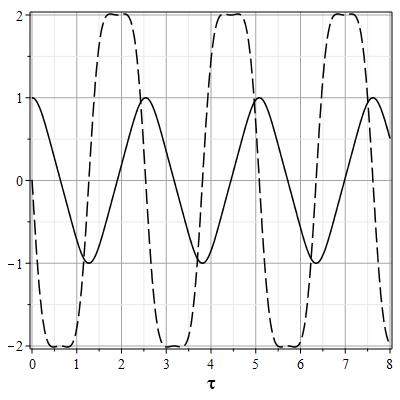}{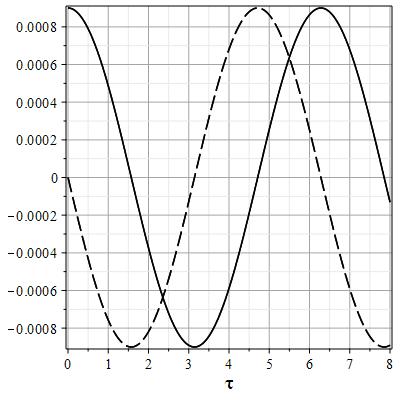}{Free oscillations of a classical scalar field: the solid curves represent $\Phi (\tau )$, and the dashed curves represent $Z(\tau )$.}{Free oscillations of a phantom scalar field: the solid curves represent $\varphi (\tau )$, and the dashed curves represent $z(\tau )$.}

\section{Differential parameters of the convergence of curves for the case of solitary scalar fields}

Let us determine the angle $\psi $ (Fig. 3) between tangent converging lines of phase trajectory~$L_{1} $ and the zero energy lines $L_{0} $ (i.e., the phase trajectory of free oscillations (Eq. (7)).  If upon convergence of the lines $L_{1} $ and $L_{0} $ this angle tends unboundedly to zero, then it can be asserted that phase trajectory \eqref{Eq2} will be tangent to the trajectory of free oscillations.  In order for phase trajectory \eqref{Eq2} to converge to the phase trajectory of free oscillations \eqref{Eq7}, it is still necessary in this case for phase velocity $v_{1} $ to remain nonzero at any point along the trajectory.

\Fig{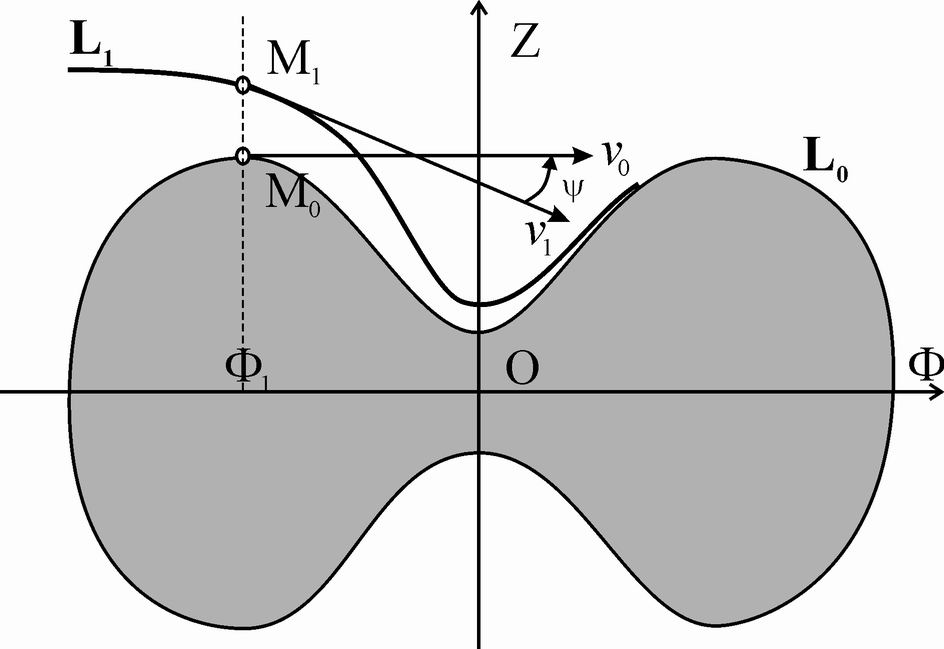}{10}{ In preparation for calculation of the differential parameters of convergence of phase trajectory $L_{1} $ to the line of zero effective energy $L_{0} $.  The points $M_{1} $ and $M_{0} $ on these lines have the coordinates $(\Phi _{1} ,Z_{1} )$ and $(\Phi _{0} ,Z_{0} )$ on phase trajectory $\Sigma _{\Phi } $.}

Let us investigate this question.  The angle $\psi $ between the two tangent vectors $v_{1} $ and $v_{0} $ is given by the well-known expression (see \cite{10}, for example)
\begin{equation} \label{Eq10}
\sin(\psi )=\displaystyle{\frac{v_{1}^{1} v_{0}^{2} -v_{1}^{2} v_{0}^{1} }{|v_{1} ||v_{0} |}}.
\end{equation}
Next, the parametric equations of the phase trajectory in the $\Sigma _{\Phi } $ plane for $\varphi =0 {\rm and} z=0$ have the form
\[L_{1} :{\rm}r(\tau )=[\Phi _{1} (\tau ),Z_{1} (\tau )],\]
where $\Phi _{1} (\tau ) Z_{1} (\tau )$ is the solution of the system of equations
\begin{equation} \label{Eq11}
\begin{array}{c} {\Phi '_{1} =Z_{1} ,} \\
 \\
 \displaystyle{{Z'_{1} =-\sqrt{3} Z_{1} \sqrt{Z_{1} {}^{2} +e\Phi _{1} {}^{2} -\frac{\alpha _{m} }{2} \Phi _{1} {}^{4} +\lambda _{m} } -e\Phi _{1} +\alpha _{m} \Phi _{1} {}^{3} .}} \end{array}
\end{equation}
As this solution, we will use the numerical solution of system \eqref{Eq5}, obtained with the help of the software package DifEqTools.  The parametric equations of the zero energy surface $\varphi =0, z=0$ take the form
\begin{equation} \label{Eq12}
\begin{array}{c} {\displaystyle{\Phi '_{0} =Z_{0} =\pm \sqrt{-e\Phi _{1} {}^{2} +\frac{\alpha _{m} }{2} \Phi _{1} {}^{4} -\lambda _{m} } ,}} \\
\\
{Z'_{0} =-e\Phi _{0} +\alpha _{m} \Phi _{0} {}^{3} .} \end{array}
\end{equation}
Here it is necessary to substitute the obtained numerical solution $\Phi _{1} $.  Thus, the tangent vectors to these curves are equal to
\begin{equation*}
\begin{array}{l}
\displaystyle{v_{1} =(\Phi '_{1} ,Z'_{1} )=(Z_{1} , -\sqrt{3} Z_{1} \sqrt{Z_{1} {}^{2} +e\Phi _{1} {}^{2} -\frac{\alpha _{m} }{2} \Phi _{1} {}^{4} +\lambda _{m} } -e\Phi _{1} +\alpha _{m} \Phi _{1} {}^{3}),} \\
\\
v_{0} =(\Phi '_{0} ,Z'_{0} )=(Z_{0} , -e\Phi _{1} +\alpha _{m} \Phi _{1} {}^{3} )
\end{array}
\end{equation*}
Thus, we shall calculate the needed differential parameters $\sin$ and $\left|V\right|$ in the case of a solitary scalar classical field.    In the case of a solitary phantom field, we find the parameters we have need of, $\sin \chi$ and $\left|v\right|$, in an analogous fashion.  In the case of the scalar doublet $\{ \Phi ,\varphi \} $ we will calculate the corresponding differential parameters in each of the corresponding planes,~$\Sigma _{\Phi }$ and $\Sigma _{\varphi } $.

Taking condition \eqref{Eq5} into account near the zero energy line, we obtain an estimate for $\sin \psi $ near this line:
\[\sin \psi \sim (-e\Phi _{1} +\alpha _{m} \Phi _{1}^{3} )(Z_{1} -Z_{0} ). \]
Hence it follows that as the curves $L_{1} $ and $L_{0} $ converge, i.e., as $\left|Z_{1} -Z_{0} \right|\to 0$, the angle between the tangents to them also tends to zero ($\psi \to 0$).

\section{Numerical modeling of the convergence of curves for the case of solitary scalar fields}

Numerical modeling of the convergence of a phase trajectory to a zero effective energy curve and calculation of the differential characteristics of this process were performed with the help of the authors' above-mentioned software package DifEqTools. In the numerical integration of system of differential equations \eqref{Eq2} we applied a 7th-to-8th order forward Runge-Kutta method, intended for integrating stiff ODE systems.  Numerous calculations confirmed the hypothesis of the existence in the investigated models of Euclidean limit cycles with effective zero energy.  Below we present particular numerical modeling examples pertaining to the type of behavior of the cosmological models which we termed \textit{adhesion} in previous papers.  In this regard, for brevity, let us consider only two cases of the parameters $P$ and the initial conditions $I$:
\begin{equation} \label{Eq13}
\mathbf{P}=[ 10,10,1,1,1,-0.1], \mathbf{I}=[0,0.4,0.1,0],\tau \in [-0.9,1.8771273]
\end{equation}
\begin{equation} \label{Eq14}
\mathbf{P}=[1,1,1,1,1,0.1],\mathbf{I}=[0,0,0.0004,0.01],\tau \in [-3.7138803,0]
\end{equation}
for a phantom field in the $\Sigma _{\varphi } $ plane.

Figures 4 and 5 show phase trajectories of these fields against the background of the forbidden regions shaded in gray. Here, a classical field at the time $\tau =-\, 0.9$ starts from a state that is far from the zero energy line, and in the final stages, for $\tau ={\rm 1.8771273}$, enters the regime of free oscillations.  A phantom field, on the other hand, at the time $\tau =-\, 3.7138803$ starts from a position very close to the zero effective energy line and then winds up in a state that is far from the zero energy line\footnote{\ It\ winds\ around\ an\ attractive\ center\ on\ the\ right-hand\ side\ of\ the\ phase\ diagram.\ }. Figures 6 and 7 display graphs of the distance of the phase trajectories from the zero energy lines.  It can be seen that the distance between these lines near the coalescence point becomes on the order of 10${}^{-10}$! Figures 8 and 9 show graphs of the sine of the angle between the tangent lines to these curves near the coalescence point.  As can be seen, the angles between the tangent lines also tend to zero, reaching values on the order of $10^{-8} -10^{-7} $. As can be seen from the graphs in Figs. 10 and 11, the velocity of motion in the phase plane at the coalescence points of the trajectories is not equal to zero, but, on the contrary, it reaches a maximum.  This is naturally explained by \textit{friction} losses experienced by the dynamical system as it enters the regime of free oscillations (see \cite{6}, for example).
\TwoFig{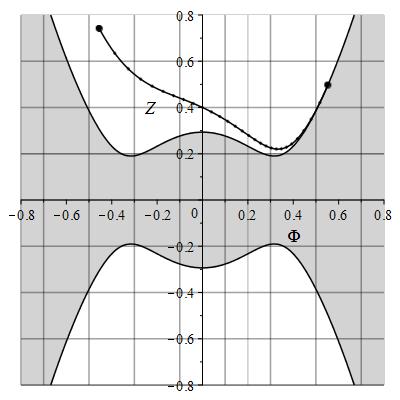}{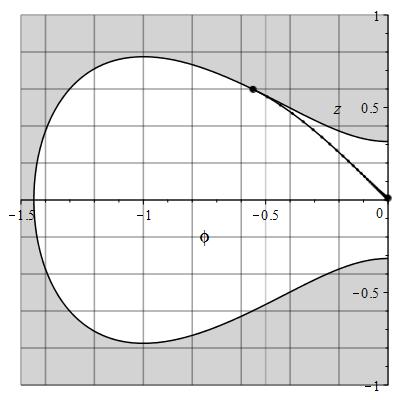}{Phase trajectory in the $\Sigma _{\Phi } $ plane for the parameters and initial conditions listed in Eqs. \eqref{Eq13}.}{Phase trajectory in the $\Sigma _{\varphi } $ plane for the parameters and initial conditions listed in Eqs. \eqref{Eq14}.}
\TwoFig{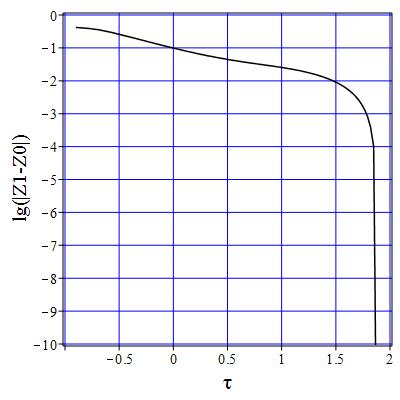}{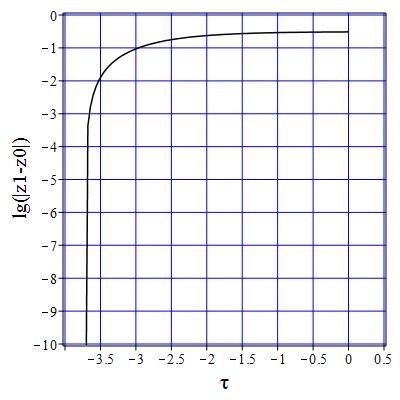}{Graph of the base-10 logarithm of $|Z_{1} -Z_{0} |$ for a classical field for the parameters and initial conditions listed in Eqs. \eqref{Eq13}.}{Graph of the base-10 logarithm of $|z_{1} -z_{0} |$ for a phantom field for the parameters and initial conditions listed in Eqs. \eqref{Eq14}.}
\TwoFig{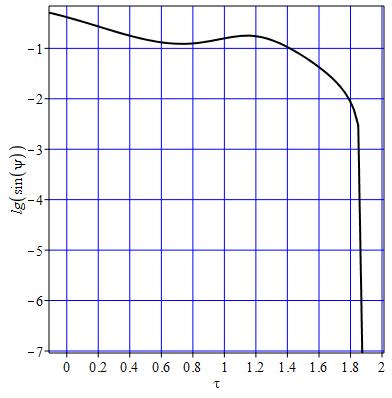}{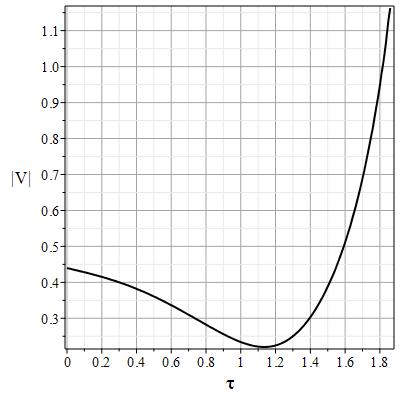}{Graph of the base-10 logarithm of $\sin \psi $ for a classical field for the parameters and initial conditions listed in Eqs. \eqref{Eq13}.}{Graphs of the base-10 logarithm of $\sin \chi $ for a phantom field for the parameters and initial conditions listed in Eqs. \eqref{Eq14}.}
\TwoFig{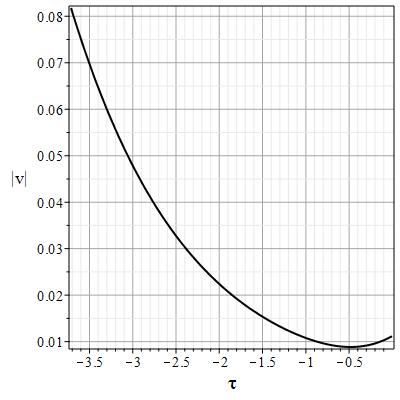}{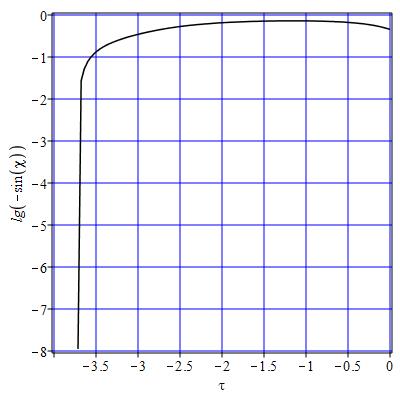}{Graph of the magnitude of the velocity $\left|V(\tau )\right|$ for a classical field for the parameters and initial conditions listed in Eqs. \eqref{Eq13}.}{Graph of the magnitude of the velocity $\left|v(\tau )\right|$ for a phantom field for the parameters and initial conditions listed in Eqs. \eqref{Eq14}.}

In conclusion, we note that we have confirmed and refined the main conclusions of \cite{3}--\cite{8} regarding the existence of Euclidean limit cycles in cosmological models based on scalar fields with a Higgs interaction potential.  In such models, the history of the Universe can contain purely Euclidean stages with a 4-dimensional Euclidean space supported by the dynamic equilibrium of oscillating scalar fields.

The work was performed within the scope of the Russian Government Program of Competitive Growth of Kazan Federal University.

\end{document}